\documentclass[5p, sort&compress]{elsarticle} 
\usepackage{graphicx}
\usepackage{amssymb}
\usepackage{epstopdf}
\usepackage[amssymb,thinqspace,thinspace]{SIunits}
\usepackage{amsmath}
\usepackage{booktabs}
\usepackage{rotating}
\usepackage{multirow}
\usepackage{psfrag}
\usepackage{adjustbox}
\usepackage[switch,modulo]{lineno}
\modulolinenumbers[5]
\usepackage{csquotes}
\usepackage{arydshln}
\usepackage{textgreek}
\usepackage{color}
\usepackage{ulem}
\usepackage[super]{nth}
\usepackage{comment}
\usepackage{lipsum}
\usepackage{dblfloatfix}
\usepackage{url}

\journal{arXiv}

\begin{document}
\begin{frontmatter}
\title{Carbon in solution and the Charpy impact performance of medium Mn steels}

\author[1]{TWJ Kwok}
\author[1,2]{FF Worsnop}
\author[1]{JO Douglas}
\author[1]{D Dye\corref{cor1}}\ead{ddye@ic.ac.uk}
\cortext[cor1]{Corresponding author}
\address[1]{Department of Materials, Royal School of Mines, Imperial College London, Prince Consort Road, London SW7 2BP, UK}
\address[2]{Department of Materials Science and Engineering, Massachusetts Institute of Technology, 77 Massachusetts Avenue, Cambridge, MA 02139, USA}

\begin{abstract}

Carbon is a well known austenite stabiliser and can be used to alter the stacking fault energy and stability against martensitic transformation in medium Mn steels, producing a range of deformation mechanisms such as the Transformation Induced Plasticity (TRIP) or combined Twinning and Transformation Induced Plasticity (TWIP $+$ TRIP) effects. However, the effect of C beyond quasi-static tensile behaviour is less well known. Therefore, two medium Mn steels with 0.2 wt\% and 0.5 wt\% C were designed to produce similar austenite fractions and stability and therefore tensile behaviour. These were processed to form lamellar and mixed equiaxed $+$ lamellar microstructures. The low C steel had a corrected Charpy impact energy (KV\textsubscript{10}) of 320 J cm\textsuperscript{-2} compared to 66 J cm\textsuperscript{-2} in the high C steel despite both having a ductility of over 35\%. Interface segregation, \textit{e.g.} of tramp elements, was investigated as a potential cause and none was found. Only a small amount of Mn rejection from partitioning was observed at the interface. The fracture surfaces were investigated and the TRIP effect was found to occur more readily in the Low C Charpy specimen. Therefore it is concluded that the use of C to promote TWIP$+$TRIP behaviour should be avoided in alloy design but the Charpy impact performance can be understood purely in terms of C in solution. 


\end{abstract}

\end{frontmatter}


\section{Introduction}

Medium Mn steels (4--12 wt\% Mn) are a relatively recent class of steels despite their conception in 1972 \cite{Miller1972}. Having been \enquote{rediscovered} as a leaner alternative to high Mn Twinning Induced Plasticity (TWIP) steels (16--30 wt\% Mn), medium Mn steels have been shown to exhibit several different plasticity enhancing mechanisms such as the Transformation Induced Plasticity (TRIP) effect \cite{Xu2022,Li2020} or a combined TWIP$+$TRIP effect \cite{Lee2014,Lee2015e}. Both mechanisms can be tailored through heat treatments and alloying to vary the strain hardening rate, leading to large elongations to failure of over 50\% \cite{Sohn2017a,Kwok2022e}. These tensile properties make medium Mn steels very suitable materials for energy absorbing applications such as automotive crash pillars \cite{Savic2018,Krizan2019}.

Current safety related automotive steels are designed to be either anti-intrusion or to crumple and absorb as much energy as possible in the event of a crash. Hot stamping or press hardening martensitic steels such as 22MnB5 are examples of anti-intrusion steels which were designed to be very strong and resist deformation \cite{Taylor2018,Fan2009}. Energy absorbing steels such as Dual Phase (DP) steels \cite{Olsson2006} are softer but significantly more ductile to allow the steel to crumple and fold, absorbing energy in the process. The opportunity for medium Mn steels, therefore, is to replace DP steels in the automotive Body in White (BIW) \cite{Savic2018,Krizan2019} as they have equivalent or better tensile properties and are also potentially cheaper due to the omission of expensive alloying elements such as Cr, Nb and V.  

The ability to exhibit the TWIP$+$TRIP effect upon deformation, therefore, was of considerable academic interest due to the prospect of activating two powerful plasticity enhancing mechanisms. Typically, TWIP$+$TRIP-type medium Mn steels do indeed exhibit larger elongations to failure compared to TRIP-type medium Mn steels ($\geq 50\%$ vs. $\geq 25\%$) \cite{Lee2014,Hu2017a,Ma2017}. The activation of the TWIP$+$TRIP effect depends on the control of Stacking Fault Energy (SFE) and stability against transformation of the austenite phase in medium Mn steels. In order to raise the SFE into the twinning regime, a large amount of C, typically more than 0.4 wt\%, is needed while keeping the Mn content within the \enquote{medium} range of between 3--12 wt\%. However, our previous work \cite{Kwok2022e} and the results by Lee \textit{et al.} \cite{Lee2014} showed that the strengthening effect from twinning was very small compared to the TRIP effect. It was therefore postulated that the large elongation in TWIP$+$TRIP-type medium Mn steels came from a very controlled TRIP effect due to the very stable and C-enriched austenite. 

Nevertheless, regardless of the strengthening contribution from TWIP or TRIP, TWIP$+$TRIP-type medium Mn steels still have higher strengths (due to the higher C content) and elongations than most TRIP-type medium Mn steels \cite{Kwok2022e}. Since the energy absorbed during plastic deformation is equal to the area under a tensile curve, it should also follow that TWIP$+$TRIP-type medium Mn steels would be more suitable for energy absorbing applications than TRIP-type medium Mn steels. Furthermore, the TWIP effect was also shown to be active at high strain rates up to approximately 2000 s$^{-1}$ \cite{Rahman2014}, while the TRIP effect is diminished at high strain rates due to adiabatic heating \cite{Rana2018}. Therefore, it is possible that the TWIP effect might begin to play a significant role at higher strain rates.



High strain rate tests such as the Hopkinson pressure bar test would be able to provide very useful information but are relatively difficult to perform \cite{Rahman2014}. Alternatively, Charpy V-notch tests can also provide some insights into the failure mechanisms, tear resistance, notch toughness and energy absorption at high strain rates of up to $10^3$ s$^{-1}$ depending on the type of material \cite{Lucon2016}. In this study, the Charpy energies of two different medium Mn steels will be compared: a high C TWIP$+$TRIP-type medium Mn steel with a mixed equiaxed $+$ lamellar microstructure, developed in previous work \cite{Kwok2022e}, and a novel low C TRIP-type medium Mn steel with a fully lamellar microstructure. This study aims to identify and compare the failure mechanisms in both steels in order to guide future alloy design. 





\section{Experimental}

Two steel ingots, High C and Low C, were vacuum arc melted using pure elements and cast into ingots measuring approximately 75 mm $\times$ 23 mm $\times$ 23 mm. The compositions of both steels as measured using Inductively Coupled Plasma (ICP) and Inert Gas Fusion (IGF) are shown in Table \ref{tab:ICPcomp}. Both steels were then homogenised at 1250 \degree C for 2 h in a vacuum furnace. The High C steel was hot rolled from 23 mm $\rightarrow$ 4 mm in thickness between 1000 \degree C and 850 \degree C in 8 passes at approximately 20\% reduction per pass. The rolled plate was quenched immediately after the last pass and subsequently intercritically annealed at 750 \degree C for 20 min. The Low C steel was hot rolled from 23 mm $\rightarrow$ 6 mm in thickness between 1000 \degree C and 950 \degree C in 6 passes at approximately 20\% reduction per pass. After the last pass, the Low C steel was returned to the furnace at 600 \degree C for 30 min then allowed to furnace cool to room temperature to simulate a coiling cycle. The Low C steel was then cold rolled from 6 mm $\rightarrow$ 4 mm. After cold rolling, the Low C steel was reheated to 950 \degree C for 5 min, water quenched and then intercritically annealed at 680 \degree C for 24 h (two-step heat treatment as described by Steineder \textit{et al.} \cite{Steineder2017}).

For comparison with strip properties, another ingot of Low C steel was rolled between 1000 \degree C and 950 \degree C from 10 mm $\rightarrow$ 3 mm in 5 passes at approximately 20\% reduction per pass. The strip was then cold rolled from 3 mm $\rightarrow$ 2 mm and heat treated in a similar manner. The final thickness of the Low C strip after descaling was 1.5 mm. The method to produce 1.5 mm strip of High C steel is described in previous work \cite{Kwok2022e}.

Subsized quarter thickness Charpy V notch samples (55 mm $\times$ 10 mm $\times$ 2.5 mm) were obtained from both High C and Low C plates in the L-T orientation (notch facing the plate transverse direction).  Three Charpy samples were tested at $-196$ \degree C, $-40$ \degree C and 22 \degree C each. Tensile samples with gauge dimensions of 19 mm $\times$ 1.5 mm $\times$ 1.5 mm were obtained from the High C and Low C plates and sheets using Electrical Discharge Machining (EDM) such that the tensile axis was parallel to the gauge length. Tensile testing was conducted using a nominal strain rate of 10$^{-3}$ s\textsuperscript{-1}. Applied strain was measured with a clip-on extensometer up to 10\% and the crosshead displacement thereafter. 

Electron Backscattered Diffraction (EBSD) and Secondary Electron Microscopy (SEM) was conducted on a Zeiss Sigma FEG-SEM with a Bruker EBSD detector. For EBSD, a 750 nm step size, dwell time of 10--15 ms and an accelerating voltage of 20 kV were used to reduce the amount of unindexed patterns below 1\%. The Bruker ESPRIT software was used to analyse the results. Secondary Electron (SE) imaging was conducted using an accelerating voltage of 5 kV.

Atom Probe Tomography (APT) specimens were fabricated using site-specific Focused Ion Beam (FIB) liftout in a Thermo Fisher Scientific Helios 5 CX DualBeam microscope from regions that contained austenite/ferrite phase interfaces at a Prior Austenite Grain Boundary (PAGB) identified using EBSD \cite{Thompson2007}. Micron scale regions of material were then mounted onto pre-fabricated silicon posts (Cameca) and were cross sectioned to identify the specific nanoscale boundaries of interest \cite{Jenkins2020}. SEM high resolution imaging at 2 kV and 0.1 nA using immersion mode was used to guide the subsequent milling. 30 kV Ga$^+$ annular milling was then used to create needle shaped specimens that contained such an interface close to the apex and the sample was then polished using 5 kV Ga$^+$ ions prior to APT analysis. 

APT analyses were carried out using a Cameca LEAP 5000 XR atom probe between a base temperature of 50 and 55 K in voltage pulsing mode with a pulse frequency of 200 kHz, a pulse fraction of 20\% and detection rates between 0.2 and 0.5\%. Data acquired was analysed using the Integrated Visualization and Analysis Software (IVAS) in AP Suite 6.1 (Cameca). Peak overlaps such as Al$^+$/ Fe$^{2+}$ at 27 Da and Si$^+$/ Fe$^{2+}$ at 28 Da were resolved based on isotopic ratios. In this study, it is acknowledged that while the APT samples were lifted from a PAGB, any interfaces identified within the analysed volume cannot be guaranteed with absolute certainty to be a PAGB without conducting further analysis, \textit{e.g.} Transmission Kikuchi Diffraction (TKD) to confirm orientation relationships. 



\begin{figure}[t!]
	\centering
	\includegraphics[width=\linewidth]{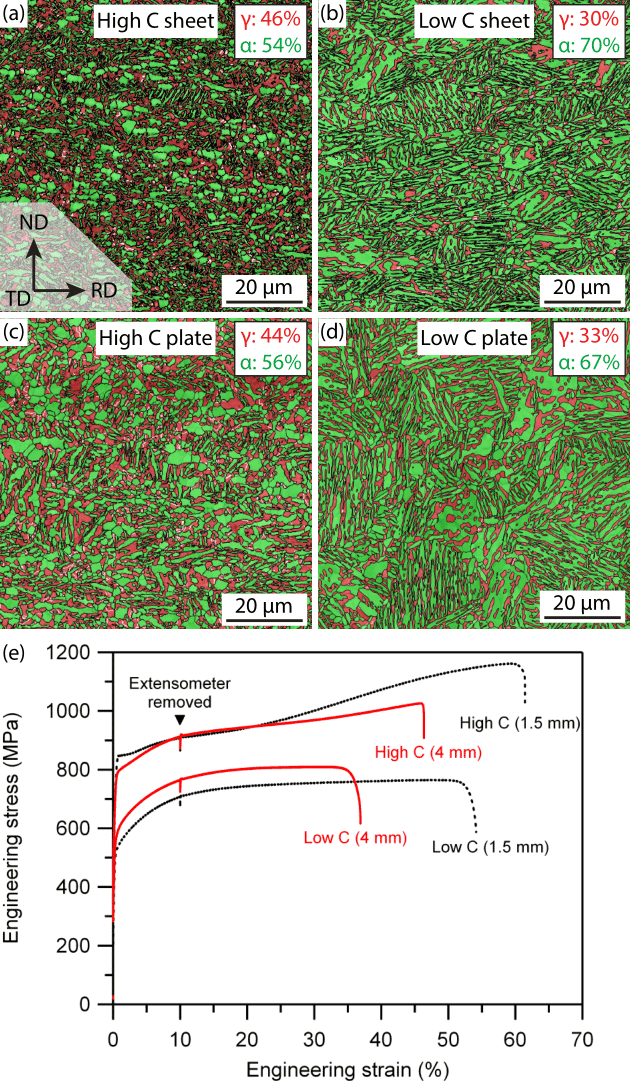}
	\caption{EBSD phase map $+$ image quality maps of (a) High C sheet, (b) Low C sheet, (c) High C plate, (d) Low C plate. Red -- austenite, green -- ferrite, phase fractions given to the nearest \%. Black lines indicate high angle grain boundaries and white lines indicate austenite $\Sigma3$ boundaries. (e) Tensile behaviour of High C and Low C sheet (dotted lines) and plate (red lines) material.}
	\label{fig:trio-quad-plate-strip-ebsd}
\end{figure}

\begin{figure*}[t!]
	\centering
	\includegraphics[width=\linewidth]{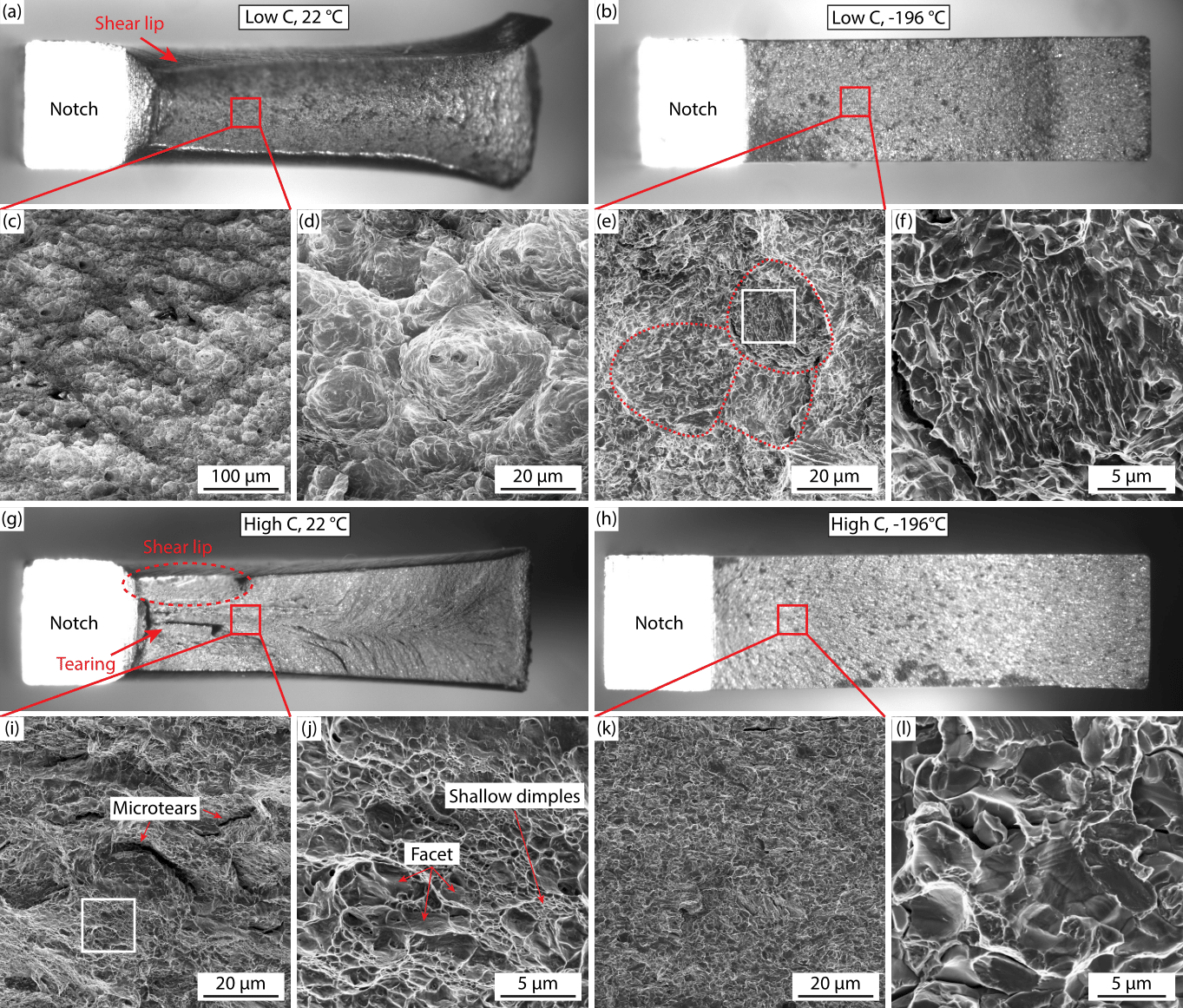}
	\caption{Stereomicrographs of postmortem Charpy samples of Low C steel tested at (a) 22 \degree C and (b) $-196$ \degree C. SE micrographs of fracture surfaces of (c) Low C, 22 \degree C, showing dimples and ductile failure in the \enquote{valley} between the adjacent shear lips, (d) magnified view of (c), (e) Low C, $-196$ \degree C, showing cleavage facets, likely PAGBs circled in dotted red lines and (f) magnified view of white square in (e) showing fracture of individual lamellae. Stereomicrographs of postmortem Charpy samples of High C steel tested at (g) 22 \degree C and (h) $-196$ \degree C. SE micrographs of fracture surfaces of (i) High C, 22 \degree C, showing microtears and (j) magnified view of the white square in (i), (k) High C, $-196$ \degree C, showing smooth facets and (l) magnified view of (k).}
	\label{fig:charpy-sem-and-macro}
\end{figure*}

\begin{table}[t]
	\centering
	\caption{Composition of the ingots used to produce High C and Low C plate steels in mass percent obtained using ICP; and IGF for elements marked with $\dagger$.}
	
	\begin{adjustbox}{width=\columnwidth,center}
		\begin{tabular}{lcccccccc}
			\toprule
			& Mn & Al & Si & C$^\dagger$ & N$^\dagger$ & S$^\dagger$ & P     & Fe \\
			\midrule
			High C & 4.35  & 3.03  & 1.46  & 0.491 & 0.003 & 0.002 & $<$0.005 & Bal. \\
			Low C & 6.30  & 2.17  & 0.99  & 0.223 & 0.004 & 0.001 & $<$0.005 & Bal. \\
			\bottomrule
		\end{tabular}%
	\end{adjustbox}
	\label{tab:ICPcomp}%
\end{table}%

\section{Results}

\subsection{Alloy design concept}

The High C steel developed in previous work \cite{Kwok2022e} had a relatively low Mn content and therefore relied on a high C content as an alternate austenite stabiliser. The high C content was able to lower the Mn content needed to create a fully austenitic hot working temperature window, raise the SFE of the austenite into the TWIP$+$TRIP regime \cite{Dumay2008,Kwok2022b} and slow TRIP kinetics \cite{Kwok2022e}. 

Therefore, when the C content in the Low C steel was reduced, the Mn content had to be increased in order to stabilise a sufficiently large austenitic hot working temperature window and raise the SFE into the lower end of the TWIP$+$TRIP regime \cite{Kwok2022b}. With the increase in Mn content, Al and Si content can also be reduced, both of which have an effect of balancing the SFE of the austenite phase \cite{Dumay2008}. A fully lamellar microstructure, rather than a mixed equiaxed$+$lamellar microstructure in the High C steel, was chosen for the Low C steel as it was reported that soft polygonal ferrite was detrimental for hole expansion in multi-phase steels \cite{Rana2016}. A fully lamellar microstructure would help to reduce any differences in strength between polygonal and lamellar ferrite grains. 

\subsection{Charpy impact testing and characterisation}

Figure 1 shows the microstructures of both plate and sheet material from High C and Low C steels. The High C steel was processed in a manner to produced a mixed equiaxed $+$ lamellar microstructure comprising of both equiaxed and lamellar ferrite with lamellar austentie grains \cite{Kwok2022e}. On the other hand, the Low C steel was processed to produce a fully lamellar microstructure. When comparing between plate and sheet microstructures, it can be seen that the overall phase fractions and microstructure morphologies were preserved. However, the grain size of the equiaxed ferrite and lamellar thicknesses were generally observed to be larger in the plate material of both steels. The larger grain size could be attributed to a combination of lower hot rolling reduction ratio per pass and slower cooling rate in the plate material.

When comparing the tensile properties in Figure \ref{fig:trio-quad-plate-strip-ebsd}e, the plate material from both steels had a lower elongation to failure. The reduced elongation may be attributed to a larger prior austenite grain size in the plate material compared to the sheet. Yield strength of the plate material was preserved within $\pm 10\%$ of the sheet material. Deformation behaviour of Low C plate was nearly identical to Low C sheet but some deviation was observed in High C plate compared to High C sheet. The deviation could be attributed to grain size effects which may affect the austenite stability and therefore the TRIP response in medium Mn steels. Nevertheless, while not perfectly identical, the plate versions of High C and Low C steels capture the essence of the tensile behaviour of their sheet counterparts.



The Charpy impact energies from both High C and Low C steels are shown in Table \ref{tab:Charpy_energy}. The Charpy impact energy from the 2.5 mm thick subsized samples (KV\textsubscript{2.5}) in J were normalised (J cm\textsuperscript{-2}) by dividing the impact energy by the cross section area of the fracture surface, \textit{i.e.} 0.8 cm $\times$ 2.5 cm. The Charpy impact energy from the 2.5 mm thick subsized sample can also be corrected to obtain the theoretical impact energy from a full sized 10 mm Charpy impact sample (KV\textsubscript{10}) using the correction method by Wallin \cite{Wallin2001,Chao2007}:

\begin{equation}
\begin{aligned}
&\frac{\text{KV\textsubscript{2.5}} \times 10}{\text{KV\textsubscript{10}} \times 2.5} = 1 - \frac{0.5 e^f}{1 + e^f} \\
&\\
\text{where} \; &f = \frac{2(\text{KV\textsubscript{10}}/2.5 - 44.7)}{17.3}\\
\end{aligned}
\end{equation}

The theoretical full sized Charpy impact energy can then be normalised by dividing the impact energy by the fracture surface, \textit{i.e.} 0.8 cm $\times$ 1.0 cm. It should be noted that there is a strong deviation from linearity between KV\textsubscript{2.5} and KV\textsubscript{10}, \textit{i.e.} 4$\times$KV\textsubscript{2.5}  $\approx$ KV\textsubscript{10}, when normalised KV\textsubscript{2.5} exceeds approximately 100 J cm\textsuperscript{-2}. This is likely to account for the increasing size of the shear lip which begins to form at higher Charpy impact energies \cite{Wallin2001}. It is acknowledged that such normalisation and correction methods are not foolproof and should be interpreted qualitatively. Quantitative comparisons should only be made with other KV\textsubscript{2.5} results in the literature.

Figure \ref{fig:charpy-sem-and-macro} shows the fracture surfaces of the post mortem Charpy impact samples under stereo-optical microscopy and SE imaging. For brevity, the Charpy samples will henceforth be referred to either Low or High C, followed by the test temperature. In Figure \ref{fig:charpy-sem-and-macro}a, the Low C, 22 \degree C sample showed very prominent shear lips on the top and bottom edges indicating ductile failure. SE micrographs in Figures \ref{fig:charpy-sem-and-macro}c-d also show a cup-and-cone type fracture surface, indicative of ductile failure. This is in contrast to the High C, 22 \degree C sample (Figure \ref{fig:charpy-sem-and-macro}g) where there was only a very small shear lip just behind the notch. SE microscopy in Figures \ref{fig:charpy-sem-and-macro}i-j show micro tears in the fracture surface as well as a mixed ductile/brittle failure mode. Several facets were observed which indicate brittle fracture but also shallow dimples which show a limited degree of ductile failure.

\begin{figure*}[t!]
	\centering
	\includegraphics[width=\linewidth]{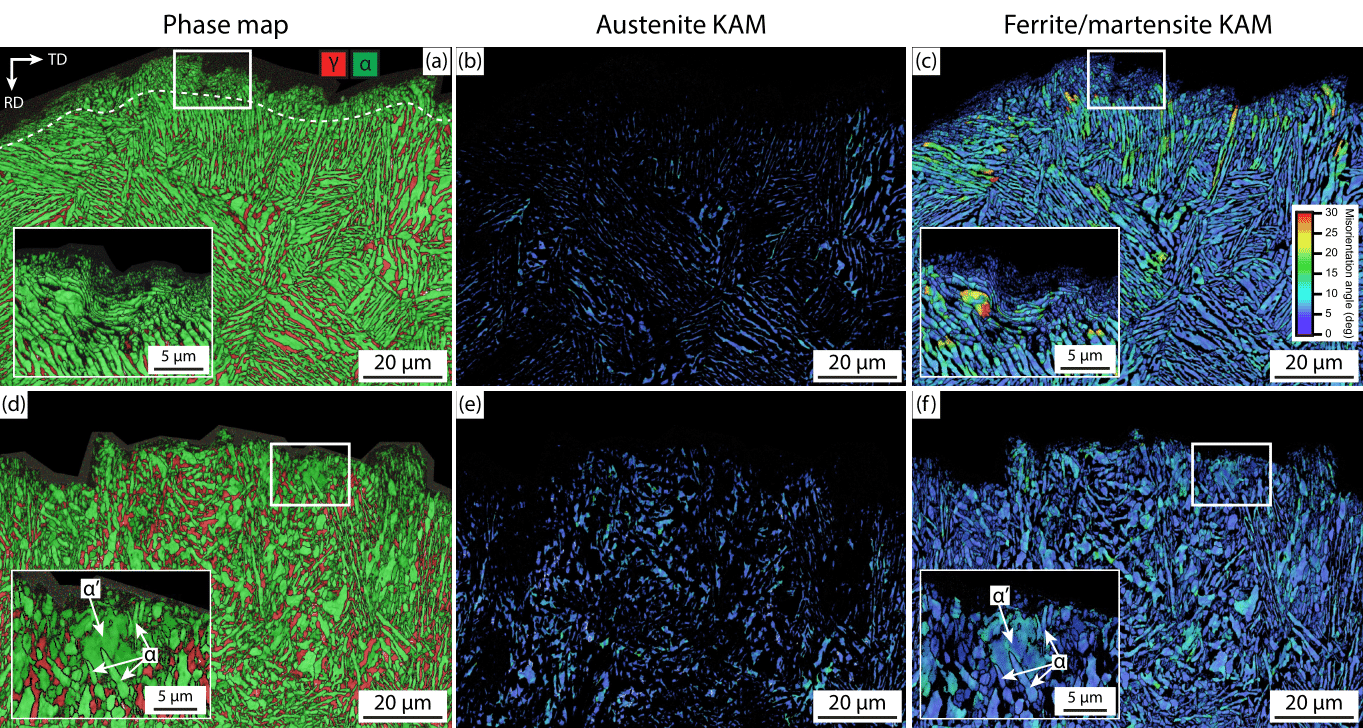}
	\caption{EBSD maps of the post mortem room temperature Charpy samples. N.B. Notch is towards the left of the micrograph and crack propagation direction is parallel to the TD. EBSD (a) Image Quality and Phase Map (IQ$+$PM), dotted line indicates the region where the austenite has almost fully transformed, (b) austenite KAM and (c) ferrite/martensite KAM maps of the fracture edge in Low C steel.  EBSD (d) IQ$+$PM, (e) austenite KAM and (f) ferrite/martensite KAM maps of the fracture edge in High C steel. Insets are high magnification maps of the respective areas bounded by the white box.}
	\label{fig:ebsd-crack-edge}
\end{figure*}

The Charpy impact samples tested at cryogenic temperatures showed brittle failure with a faceted fracture surface regardless of composition. In the Low C, $-196$ \degree C sample (Figures \ref{fig:charpy-sem-and-macro}b, e-f), the facets had a corrugated appearance which may correlate with the lamellar microstructure within a prior austenite grain. This might suggest that the crack may have propagated through a prior austenite grain, rather than along the PAGBs as shown by Han \textit{et al.} \cite{Han2017} in a Fe-7Mn-0.5Si-0.1C steel. In the High C, $-196$ \degree C sample, the facets were very smooth and equiaxed in morphology.

\begin{table}[t!]
	\centering
	\caption{Normalised and corrected Charpy impact energies (KV\textsubscript{B}, where B is the thickness of the Charpy impact sample) from High C and Low C steels tested at various temperatures. Standard errors in parantheses. N.B. Normalised -- Charpy energy divided by cross sectional area. Corrected -- KV\textsubscript{2.5} converted to KV\textsubscript{10}. *Corr+norm--corrected and normalised.}
	\begin{adjustbox}{width=\columnwidth,center}
		\begin{tabular}{lccc}
			\toprule
			& $22$ \degree C    & $-40$ \degree C   & $-196$ \degree C \\
			\midrule
			\textit{High C}	&&&\\
			KV$_{2.5}$ (J) & 12.9 (0.7) & 4.3 (0.3) & 1.6 (0.3) \\
			Normalised KV$_{2.5}$ (J cm$^{-2}$) &	64.5 (4)	&	21.5 (1.5)	&	8.0 (1.5)\\
			Corrrected KV$_{10}$ (J) & 53    & 17    & 6 \\
			\smallskip
			Corr$+$norm KV$_{10}$ (J cm$^{-2}$) & 66    & 21    & 8 \\
			\textit{Low C} &&&\\
			KV$_{2.5}$ (J) & 31.9 (0.7) & NA    & 0.9 (0.4) \\
			Normalised KV$_{2.5}$ (J cm$^{-2}$) & 159.5 (3.5) & NA    & 4.5 (2.0) \\
			Corrected KV$_{10}$ (J) & 256    & NA    & 4 \\
			Corr+norm KV$_{10}$ (J cm$^{-2}$) & 320   & NA    & 5 \\
			\bottomrule
		\end{tabular}%
		\label{tab:Charpy_energy}%
	\end{adjustbox}
\end{table}%

EBSD maps were obtained from the post mortem room temperature Charpy samples and shown in Figure \ref{fig:ebsd-crack-edge}. In the Low C, 22 \degree C sample, the phase map (Figure \ref{fig:ebsd-crack-edge}a) showed a thin region of approximate 5--10 \textmu m thick below the fracture surface where austenite was not detected. This strongly suggests that the austenite within this region (TRIP zone) had completely transformed into martensite, indicating that the TRIP effect was active. It cannot be said definitively if the crack propagated along the PAGBs or across the prior austenite grains. When comparing the austenite and ferrite/martensite Kernal Average Misorientation (KAM) maps in Figures \ref{fig:ebsd-crack-edge}b and c respectively, it was observed that the austenite lamellae were hardly deformed, while the ferrite lamellae were plastically deformed at a significantly greater depth than the TRIP zone.

In the High C, 22 \degree C sample, the phase map in Figure \ref{fig:ebsd-crack-edge}d did not show the same uniform subsurface TRIP zone and austenite was still observed very close to the fracture surface. However, the TRIP effect was still active in this sample. In Figure \ref{fig:ebsd-crack-edge}d, martensite can be qualitatively identified based on its larger blocky morphology and lower indexing quality (appears darker) compared to the surrounding ferrite. Therefore, martensite could be identified immediately below the fracture surface in the High C, 22 \degree C sample. 

\begin{figure}[t!]
	\centering
	\includegraphics[width=\linewidth]{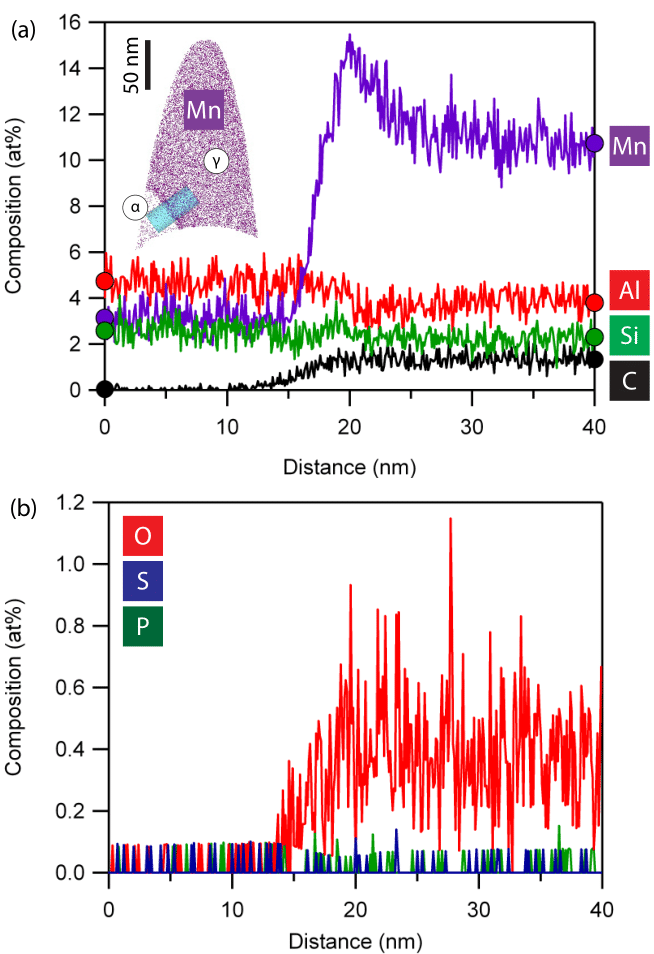}
	\caption{APT results obtained from a needle containing a PAGB in the Low C plate steel. (a) Concentration profile across an $\gamma/\alpha$ interface. Full circles at 0 nm and 40 nm indicate the far-field composition of ferrite and austenite respectively. Inset: Mn atom map and location of cylinder used to measure the concentration profile within the needle. (b) Concentration profile of tramp elements within the same volume as (a). }
	\label{fig:quadrolloy-apt}
\end{figure}

Figure \ref{fig:quadrolloy-apt} shows the results from an APT needle obtained from a PAGB in the Low C plate steel. Unfortunately, the interface from the APT needle obtained from the High C plate steel was lost due to a microfracture event during analysis but compositions from both phases could still be obtained. The compositions of austenite and ferrite phases from both needles, obtained \textit{via} APT, are shown in Table \ref{tab:APTcomp}. SFE was determined from the austenite compositions using the method by Sun \textit{et al.} \cite{Sun2018}. The martensite start (Ms) temperature was determined using the equation by Kaar \textit{et al.} \cite{Kaar2021} and the Md\textsubscript{30} temperature was determined using the equation by Angel \cite{Angel1954} and Nohara \textit{et al.} \cite{Nohara1977}. From Table \ref{tab:APTcomp}, the austenite phase in both Low C and High C steels have SFE ranges within the TWIP$+$TRIP regime in medium Mn steels \cite{Kwok2022b}. However, the austenite phase in the High C steel was significantly more stable against transformation as seen by the lower Md\textsubscript{30} temperature.

From Figure \ref{fig:quadrolloy-apt}a, the austenite phase could be identified as the Mn and C enriched phase, while the ferrite phase could be identified as the Mn and C depleted but Al enriched phase. It is noteworthy that Si was not observed to partition strongly to either phase although a slight enrichment of Si was observed at the interface. This Si enrichment at the interface was also observed in previous work \cite{Kwok2019}. The Mn profile in the ferrite phase was relatively constant but in the austenite phase, the Mn content appeared to decrease with distance away from the interface moving into the grain interior. This was likely due to the sluggish diffusion of Mn in FCC austenite under Partitioning Local Equilibrium (PLE) mode \cite{Ding2018}, \textit{i.e.} Mn at the interface struggles to diffuse into the austenite interior. 

Figure \ref{fig:quadrolloy-apt}b shows the concentration of tramp elements O, S and P within the same sampled region as Figure 2\ref{fig:quadrolloy-apt}. It could be observed that the austenite phase had a greater solubility for O. However, there was no segregation of tramp elements to the grain boundary. Elements such as B and N were not detected above noise background levels and therefore omitted from Figure \ref{fig:quadrolloy-apt}b. Therefore, from Figure \ref{fig:quadrolloy-apt}, there was no significant segregation of solute, interstitial nor tramp elements to the PAGB within the APT needle obtained from the Low C plate steel.

\begin{table}[t!]
	\centering
	\caption{APT composition analysis of the High C and Low C plate steels in wt\%. $\dagger$ C content determined by lever rule using C content measured by IGF and phase fractions obtained using EBSD, assuming negligible C content in ferrite. N.B. A lower Md\textsubscript{30} temperature indicates a more stable austentite against deformation induced transformation. }
	\small
	\begin{adjustbox}{width=\columnwidth,center}
		\begin{tabular}{lcccccccc}
			\toprule
			& Mn    & Al    & Si    & C$^\dagger$    & SFE   &  Ms  &  Md\textsubscript{30}	\\
			& \multicolumn{4}{c}{}   & (mJ m$^{-2}$) & (\degree C) & (\degree C) \\
			\midrule
			High C $\gamma$  & 7.4 & 3.0 & 1.3 &  1.12   & 41  & -53 & -19 \\
			\smallskip
			Low C $\gamma$  & 11.0 & 1.9 & 1.2 & 0.68   & 25  & -80 & 155	 \\
			
			High C $\alpha$   & 2.3 & 3.4 & 1.8 & --  & --  & -- & -- \\
			Low C $\alpha$   & 3.2 & 2.4 & 1.4 & --  &  -- & -- & -- \\
			
			\bottomrule
		\end{tabular}%
	\end{adjustbox}
	\label{tab:APTcomp}%
\end{table}%

\section{Discussion}


Perhaps the most interesting result from this comparative study was that the Charpy impact energy of the Low C steel was significantly larger than the High C steel, despite having a lower yield strength, tensile strength and elongation (Figure \ref{fig:trio-quad-plate-strip-ebsd} and Table \ref{tab:Charpy_energy}). While higher yield strengths tend to correlate with improved energy absorption in drop tower crush tests \cite{Link2017,Ratte2006}, it seems that the same correlation does not exist between tensile properties and Charpy impact performance \cite{Zaera2014}. Of the total energy absorbed in a Charpy impact test, Sugimoto \textit{et al.} \cite{Sugimoto2015} found that the energy expended to initiate a crack was relatively constant in medium Mn steels, regardless of strength, Mn content or volume fraction of austenite. Instead, it was the energy expended to propagate the crack which dominated the total energy absorption. Therefore, any crack retarding or blunting mechanisms in the steel will be expected to greatly improve the Charpy impact performance of the steel. 

\subsection{Effects of microstructure}

\begin{figure}[t]
	\centering
	\includegraphics[width=\linewidth]{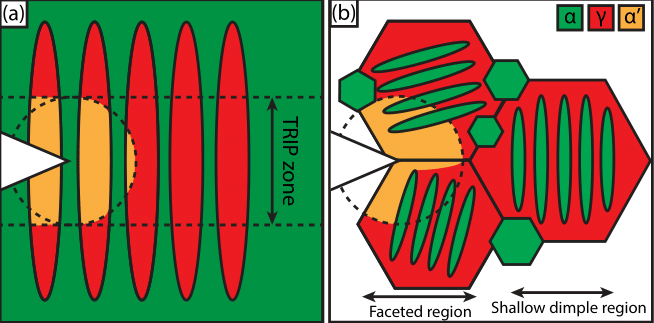}
	\caption{Schematic of martensite transformation with crack propagation in (a) Low C within a PAG and (b) High C steel across several PAGs with different ferrite lamella orientations. Stress field ahead of the crack tip indicated by the dotted circle. N.B. the PAG in the Low C steel is ferritic with austenite lamella and the other way around in the High C steel. }
	\label{fig:cracking-schematic}
\end{figure}

In the Low C, 22 \degree C sample, a 5 -- 10 \textmu m TRIP zone was observed immediately beneath the fracture surface. The austenite below the TRIP zone did not appear to be significantly deformed (Figures \ref{fig:ebsd-crack-edge}a--c). This suggests that the austenite transformed to martensite rather than undergo plastic deformation under the stress at the crack tip. The microstructure ahead of the crack tip would then resemble a laminate composite comprising of alternating layers of soft ferrite reinforced by layers of hard martensite. Cao \textit{et al.} \cite{Cao2017} showed that ultrahigh Charpy impact energies ($>$400 J cm\textsuperscript{-2}) could be obtained in a steel with a ferrite/martensite laminated microstructure. The softer and more ductile ferrite lamella were also able to transmit the stress deeper into the material which explains why the tips of the ferrite lamellae away from the fracture surface also experienced significant plastic strain (Figure \ref{fig:trio-quad-plate-strip-ebsd}c). A schematic of the described process in the Low C, 22 \degree C sample is shown in Figure \ref{fig:cracking-schematic}a.

Therefore, the energy expended during crack propagation in the Low C steel was used to transform the austenite to martensite within the TRIP zone, tear through a ferrite/martensite laminate structure and deform the surrounding ferrite lamella far away from the TRIP zone. The austenite to martensite transformation in itself does not absorb significant amounts of energy \cite{He2019} but Song \textit{et al.} \cite{Song2000} suggested that the transformation helped relax the stress at the crack tip suppressing void formation. Nevertheless, a significant amount of energy expended during crack propagation in the Low C steel was used to tear through the laminate structure.



In the High C, 22 \degree C sample, the stress ahead of the crack tip would similarly cause the austenite to transform to martensite. However, due to the mixed morphology of the ferrite phase and also a higher austenite fraction, there may not always be bridging ferrite lamella to blunt the crack tip. Therefore, large uninterrupted regions of austenite could transform into martensite which might subsequently cleave open.  Austenite grains are also not always kept seperate from each other, implying a large amount of $\gamma/\gamma$ grain boundaries. If a $\gamma/\gamma$ boundary was caught in the stress field, it will turn into a $\alpha'/\alpha'$ boundary after transformation which might also cleave open. Both of these factors may result in the crack being able to propagate rapidly through the microstructure. Where the crack was able to propagate rapidly, there would likely be very little subsurface plastic deformation \textit{i.e.} stress shielding \cite{Zhang2013}, as observed in Figures \ref{fig:ebsd-crack-edge}e-f. In certain areas, even the austenite grains just below the fracture surface were protected from transformation. A schematic of the described mechanism in the High C steel is shown in Figure \ref{fig:cracking-schematic}b. The facets observed in Figures \ref{fig:charpy-sem-and-macro}i-j could therefore correspond to the cleavage surfaces of the martensite grains in the High C steel.

Therefore, the energy absorbed during crack propagation in the High C steel was consequently lower than the Low C steel as the crack was able to propagate \textit{via} brittle fracture of large areas of connected martensite (previously austenite) grains. This effect was coined the \enquote{brittle network} effect by Jacques \textit{et al.} \cite{Jacques2001a} who similiarly found a decrease in resistance to cracking in a steel with a larger volume of high carbon retained austenite. Future medium Mn alloy development should focus on isolating austenite grains in order to improve resistance to cracking. 

The morphology of the austenite and ferrite grains therefore appear to be a significant factor in the Charpy impact performance of medium Mn steels. Song \textit{et al.} \cite{Song2000} showed in low alloy TRIP steels that the TRIP effect was most beneficial when the austenite phase was in the form of films between bainitic laths as compared to blocky islands. However, Han \textit{et al.} \cite{Han2017} found that the room temperature Charpy impact performance was very similar between fine grained equiaxed and lamellar microstructure variants, both having the same bulk composition and austenite fraction. This suggests that microstructure may not be the only factor influencing the Charpy impact performance of medium Mn steels.





\subsection{Effects of composition and segregation}

Aside from differences in microstructure, the two investigated medium Mn steels had very different compositions with the High C steel having a greater alloy content in all major elements:  Mn, Al, Si and C. While the effects of individual elements on the Charpy impact performance have not been investigated in medium Mn steels, C was expected to be the most significant element. The ASM handbook \cite{Roe1990} showed that increasing the C content generally leads to a higher Ductile-Brittle Transition Temperature (DBTT) but a reduced upper shelf energy in various ferritic/martensitic steels. On the other hand, in fully austenitic TWIP steels, C has the effect of strengthening the austenite phase and improving the absorbed impact energy \cite{Bordone2021}. In TWIP$+$TRIP-type Fe-Cr-Mn stainless steels, Hwang \textit{et al.} \cite{Hwang2012} showed that there was no significant difference in room temperature Charpy impact energy between C contents of 0.2--0.4 wt\%.

C also significantly influences the kinetics and extent of the TRIP effect by stabilising the austenite phase, \textit{i.e.} increasing the resistance to deformation induced martensitic transformation. The austenite stability of the High C steel was consequently significantly higher than the Low C steel (Table \ref{tab:APTcomp}). While both High C and Low C exhibited the TRIP effect, it was difficult to quantify the extent of the TRIP effect just below the fracture surface. Furthermore, due to stress shielding effects in the High C steel, the extent of transformation could not be attributed to composition alone.


Nevertheless, depending on the C content, the transformed martensite will vary in hardness and therefore brittleness \cite{DelaConcepcion2015,Gao2019a}. The strength of the transformed martensite, $\sigma_{\alpha'}$ can be estimated using the equation \cite{Lee2014,Speich1968}:

\begin{equation}
\sigma_{\alpha'} \, \text{(MPa)} = 413 + 1720 \, X_C
\end{equation}

\noindent
where $X_C$ is the C content in wt\%. Based on the C content of the High C and Low C steels in Table \ref{tab:APTcomp}, $\sigma_{\alpha'}$ in the High C and Low C steel would be 2.3 GPa and 1.6 GPa respectively. The martensite in the High C steel was therefore expected to be very brittle \cite{Sun2017c}. Therefore, while a stronger martensite might be preferable for higher tensile strengths and resistance to necking from the perspective of a tensile test (Figure \ref{fig:trio-quad-plate-strip-ebsd}e), it may not be as beneficial in terms of crack resistance. 

These results therefore show that the Charpy V-notch impact properties of medium Mn steels appear to be TRIP-limited. The morphology, composition,  strength and ductility of the martensite phase heavily influence the crack propagation energy during the impact test. While not investigated in this study, the TWIP effect would therefore only be expected to play a limited role.

On the other hand, there is a growing body of literature demonstrating segregation of elements to certain interfaces such as PAGBs \cite{Han2017} or $\delta$-ferrite boundaries \cite{Kim2019a} leading to poor cohesion and reduced impact properties. APT was conducted on the Low C sample and Figure \ref{fig:quadrolloy-apt} shows a ferrite/austenite boundary in a needle lifted from a PAGB. The results do not show any concentration spike of Mn, C or any other tramp elements to the identified boundary. This gives confidence that segregation does not always occur in medium Mn steels. Segregation of elements such as Mn and C also appears to be a time-related issue. For medium Mn steels where segregation was identified \cite{Kim2019a,Han2017,Kwok2019}, the IA duration was $\leq 1$ h. In this study, the Low C steel was intercritically annealed for 24 h to replicate the batch annealing process. This suggests that batch annealed medium Mn steels might be less susceptible to segregation related embrittlement.








\section{Conclusions}

The Charpy impact properties of two different medium Mn steels with different microstructures, tensile properties and compositions were compared. Several key findings are shown below.

\begin{enumerate}
	\item Both the Low C and High C steels exhibited the TRIP effect along the fracture edge. However, the Low C steel had a significantly higher absorbed Charpy impact energy compared to the High C steel. The reasons for which could be attributed to microstructure and C content. 
	\item A lamellar microstructure absorbs more energy during crack propagation compared to a mixed equiaxed $+$ lamellar microstructure by acting as a laminate composite. The austenite within the stress field transforms into martensite and reinforces the ferrite matrix. 
	\item Austenite containing a high C content consequently transforms to high C martensite, which is strong but very brittle. Formation of high C martensite might be beneficial in a tensile test but might be deletrious in a Charpy impact test especially if the martensite grains are able to form a continuous network in the microstructure.
	\item Apart from Mn partitioning effects, segregation of solute, interstitial and tramp elements to the PAGB were not detected in the Low C steel which may be attributed to long intercritical annealing durations. 
\end{enumerate}


\section{Acknowledgements}
TWJK gratefully acknowledges the provision of a studentship from A*STAR, Singapore. We gratefully acknowledge the Engineering and Physical Science Research Council for funding the Imperial Centre for Cryo Microscopy of Materials at Imperial College London (EP/V007661/1).

\bigskip
\textbf{References}
\bigskip

\end{document}